\documentstyle[preprint,aps]{revtex}
\begin{document}
\title{Conditions for Soliton-Like Wave Propagation in\\Pockels and Photorefractive Media}
\author{S. Abbas Hosseini, $ {\rm {S.M. Mahdavi }}^*$ and  Debabrata Goswami} 
\address{Tata Institute of Fundamental Research, Homi Bhabha Road,
Mumbai 400 005, India.}
\address {$^*$Department of Physics, Sharif University of Technology, Tehran,Iran}
\date{\today}
\maketitle
\begin{abstract}
We study the conditions for  soliton-like wave propagation in the Photorefractive (PR) and electro-optic (i.e., Pockels) material, by using Nonlinear Schrodinger (NLS) equation. The complete NLS equation is solved analytically and numerically by transforming it into the phase space. Our results clearly show the existence of both the dark and bright solitary solutions for the PR medium.  Interestingly, however, we find only one bright solitary solution in the Pockels case and there is no evidence of any dark solitary solution.
\end{abstract}

\pacs{42.65.Tg, 42.25.Bs, 42.81.Dp}

\section{Introduction} 
In a uniform Kerr-like nonlinear medium, the transverse profile of an optical beam propagates unchanged since the diffraction is balanced by self-focusing. Such self-focused waves are called solitons, and are described by the Nonlinear Schrodinger (NLS) equation in space-time dimension \cite{m1} - \cite{m4} . Zakharov and Shabat have solved such NLS equation for the Kerr medium by using Inverse Scattering Transformation (IST)\cite {m2}. Self-focused optical beams in other non-Kerr, nonlinear media have also been investigated for the possibility of generating soliton like motion. For example, in the case of an optical beam propagation in a Photorefractive (PR) material, charge carriers are photo-excited and redistributed around the beam. This charge distribution produces a space charge field, which in turn modulates the refractive index(the electro-optic effect) \cite {m5},\cite {m6}. When such a system exhibits optical  nonlinearity, the related modifications in the refractive index induces either a self-focusing or self-defocusing of the incident  beam. With an appropriate choice of the experimental parameters, the input beam profile can be made to converge asymptotically to a soliton state whose PR nonlinearity compensates for the diffraction, and the beam profile remains unchanged as it propagates. Montemezzani and Gunter have investigated the profile of the one-dimensional bright spatial solitons by using a power series expansion \cite {m7}. They managed to calculate the first four terms of the series that describe the profile of the bright solitons.  However, a complete temporal and spatial solution is yet to emerge.
 
In this work, we first attempt  to solve the complete NLS equation for the Pockels medium. It is extremely difficult to solve this equation with IST, so we invoke the technique of transforming the NLS equation into phase space and determine its exact solution. We then extend the technique to solve the NLS equation for the PR medium. In fact, we find that the technique of phase space transformation is so general that it  enables us to solve the NLS equation for any polynomial potential that appear in most optics and plasma physics problems. 

\section{Solitary wave in dispersive Pockels media}

Let us consider the spatial and temporal variation of the electric field $\tilde E(z,t)$ of an optical pulse propagating through a linear electro-optic medium as\cite {m8}: 

\begin{equation}
\tilde E(z,t)=\tilde A(z,t)exp~i(k_0z-\omega _0t)+c.c.
\end{equation}

\noindent
Here $z$ and $t$ are the space and time coordinates and $k_0$ and $\omega _0$ are the linear components of the wavevector and frequency of the electric field.  The total wavevector k can be written as\cite {m8}:

\begin{equation}
k=k_0+\Delta K_{NL}+(\omega -\omega _0) (\frac{dk}{d\omega})_{\omega =\omega _0} +\frac {1}{2} (\omega -\omega _0)^2 (\frac{d^2k}{d\omega ^2})_{\omega =\omega _0}, 
\end{equation}

\noindent
the nonlinear contribution to the propagation constant is: 

\begin{equation}
\Delta K_{NL}=\gamma |E(z,t)|=\gamma |A(z,t)|
\end{equation}

\noindent
where $\gamma$ is Pockels constant. Defining the retarded time $\tau $ as $\tau =t-z/v_g$, the NLS equation can be written as \cite {m8},\cite {m9}:

\begin{equation}
{\frac{\partial \tilde A}{\partial z}}+i\frac{\beta _2}2{\frac{\partial
^2\tilde A}{\partial \tau ^2}}=i\Delta K_{NL} \tilde A,
\end{equation}

\noindent
where $\beta _2$ is a measure of group velocity  dispersion (GVD) and is
defined in terms of wavevector as: 

\begin{equation}
\beta _2=(\frac{d^2k}{d\omega ^2})_{\omega =\omega _0}=\frac d{d\omega
}(\frac 1{v_g(\omega )})_{\omega =\omega _0}=(\frac{-1}{v_g^2}\frac{dv_g}{%
d\omega })_{\omega =\omega _0}.
\end{equation}

\noindent
Using Eqn.(3), the NLS equation for the propagating pulses in a Pockels medium  is: 

\begin{equation}
{\frac{\partial \tilde A}{\partial z}}+i\frac{\beta _2}2{\frac{\partial
^2\tilde A}{\partial \tau ^2}}=i\gamma |\tilde A|\tilde A,
\end{equation}

The effect of GVD on propagation of pulse in a linear dispersive medium can
be considered by setting $\gamma =0$ in Eqn.(6), 

\begin{equation}
{\frac{\partial \tilde A}{\partial z}}+i\frac{\beta _2}2{\frac{\partial
^2\tilde A}{\partial \tau ^2}}=0.
\end{equation}

Eqn.(7) can be solved using the Fourier method \cite{m8}, which results in the familiar result that the long wavelength components of an optical pulse propagate faster than the short wavelength components when the GVD parameter, $\beta _2$, is positive, and vice versa.

Another interesting phenomenon that occurs in the nonlinear media is self-phase modulation effect (SPM) , which can be determined in the absence of dispersion ($\beta _2=0$). When the SPM parameter $\gamma$ is negative, the short wavelength components of an optical pulse propagate faster than the long wavelength components, and vice versa \cite {m8}, \cite {m9}.
So when $\beta _2>0$ and $\gamma<0 $ or $\beta _2<0$ and $\gamma>0$, the SPM effect is canceled by GVD effect, and optical pulse can propagate in a medium with an unchanged shape. Let us consider a solution to the Eqn.(6) as follows, 

\begin{equation}
A(z,\tau )=F(\tau )exp~(i\Gamma z),
\end{equation}

\noindent
where $\Gamma$ is defined as the wave propagation constant. Defining new
variables as: 

\begin{equation}
\cases {\lambda =2\frac {\Gamma} {\beta _2},\cr
\mu =-2 \frac{\gamma }{\beta _2}.\cr}
\end{equation}

\noindent
the Eqn.(6) becomes 

\begin{equation}
\frac{d^2F(\tau )}{d\tau ^2}+sgn(\lambda )|\lambda |F(\tau )+sgn(\mu )|\mu
|F^2(\tau )=0,
\end{equation}

\noindent
where sgn($\lambda $) and sgn($\mu $) takes the value of plus or minus one.

Making a further change of variables as follows: 

\begin{equation}
\cases {\sqrt{|\lambda |}\tau =T,\cr
F(\tau )=\frac{|\lambda |}{|\mu |}Q(\tau ),\cr
Q^{^{\prime }}={\frac{dQ(T)}{dT}},\cr}
\end{equation}

\noindent
Eqn.(10) reduces to: 

\begin{equation}
Q^{^{\prime \prime }}+sgn(\lambda )Q+sgn(\mu )Q^2=0.
\end{equation}

Let us consider this equation as an Euler-Lagrange equation of a Hamiltonian   
system. Multiplying Eqn.(12) by $Q^{\prime}$ and integrating with respect to
time, the corresponding Hamiltonian function can be derived as the following form:

\begin{equation}
H(Q,Q^{\prime })=Q^{\prime  2}+sgn(\lambda )Q^2+\frac {2}{3} sgn(\mu )Q^3=h,
\end{equation}

\noindent
where $h$ is the constant of integration and can be determined from the
initial conditions. The Hamiltonian in Eqn.(13) represents a dynamical system and its behavior is investigated by considering the quantities $\lambda, \mu $ and $h$ as parameters.

When the parameters  $\lambda < 0$ and $\mu > 0$, Eqn.(13) can be expressed as: 

\begin{equation}
H(Q,Q^{\prime})= Q^{\prime 2}+{\frac{2}{3}}Q^3-Q^2=h. 
\end{equation}

\noindent
We can find the extreme limits (extrema points) of the Hamiltonian \cite{m10}  
by setting \begin{equation}
 \cases {\frac{\partial h}{\partial Q^{\prime}} =0,\cr
  \frac{\partial h}{\partial Q} =0.\cr}
\end{equation}

We obtain $h=0$ and $ h=-\frac{1}{3}$ as the extrema points. Now we can plot $ Q^{\prime 2} - Q^2 + \frac{2}{3} Q^3 = h $ for different values of $h$, which are: $h=0 $, $ -\frac{1}{3} < h < 0 $ , $h > 0$ and $h = -\frac{1}{3} $
(see Fig. (1)). For $h=0$, the figure suggests a hyperbolic solution. Putting $h=0$ in Eqn. (14) we obtain the following differential :

\begin{equation}
\frac{d Q}{\sqrt{\frac{-2}{3} Q^3 + Q^2}} = \pm dT
\end{equation}

\noindent
by integration we can obtain the following solution
\begin{equation}
Q=\frac{3}{2} sech^2 (\frac{T}{2}), 
\end{equation}

\noindent
writing in terms of the original variables, we obtain

\begin{equation}
A(z,t) = \frac{3 |\lambda |}{2 |\mu |}  sech^2 (\frac{\sqrt {|\lambda |}}{2} (t - z/v_g) ) e^{i \Gamma z}. 
\end{equation}

\noindent
When $(t -  z/v_g )\rightarrow \pm \infty$,  $A(z,t) \rightarrow 0$ and hence this is a bright solitary solution,
a profile of which is shown in Fig.(2).

In the other case, $\lambda >0$ and $\mu>0$ we find: 

\begin{equation}
H(Q,Q^{^{\prime}})= Q^{^{\prime}2}+{\frac{2}{3}}Q^3+Q^2=h. 
\end{equation} 

\noindent
Calculations similar to above can be done to get the extremum point to be   $(Q,Q^{\prime})=(0,-1) $.  We plot $H(Q,Q^{\prime})$ for different values of $h$ in Fig.(3). We have hyperbolic solution for $h={\frac{1}{3}}$. For this value of $h$, Eqn. (19) can be integrated as before as: 

\begin{equation}
Q(T)= {-\frac{3}{2}}\tanh ^2({\frac{T}{2}})+{\frac{1}{2}}. 
\end{equation}

\noindent
Thus, 
\begin{equation}
A(z,t) = \frac{|\lambda|}{|\mu|} [ -\frac{3}{2} tanh^2 ( \frac{\sqrt{\lambda}}{2} (t -  z/v_g ) ) + \frac{1}{2} ] e^{i \Gamma z}
\end{equation}

The profile of the solution is given in Fig.(4). Since this amplitude assumes negative values, it is a unphysical solution and as such no dark solitons are possible. The case of $(\lambda <0 $ and $\mu<0)$ or $(\lambda >0$ and $\mu<0)$ also lead
to non-physical solution.

\section{Solitary wave in PR media}

The nonlinear contribution to the propagation in PR media can be written as (see appendix):

\begin{equation}
\Delta K_{NL}=\gamma {\frac{I_p}{I_p+I_s(z,t)}}, 
\end{equation}

\noindent
where $\gamma$ is the electronic coupling parameter, $I_p$ is the intensity of the pump beam and $I_s(z,t)$ is the spatial and temporal variation of the signal beam. When this form of the nonlinear contribution is substituted into the NLS Eqn.(4), we obtain: 

\begin{equation}
{\frac{\partial \tilde A}{\partial z}}+i\frac{\beta _2}2{\frac{\partial
^2\tilde A}{\partial \tau ^2}}=i\gamma {\frac{I_p}{I_p+I_s}}\tilde A, 
\end{equation}

\noindent
where $\beta _2$ is a measure of the dispersion of the group velocity.

\noindent
To achieve the separation of variables $z$ and  $\tau$, we seek a solution of the form: 

\begin{equation}
A(z,\tau )=\sqrt{I_p}F(\tau )exp~(i\Gamma z), 
\end{equation}

\noindent
where $\Gamma $ is defined as the pulse propagation constant and $F(\tau )$
is a real function of $\tau $. With the definition: 

\begin{equation}
\cases {\lambda =2\frac {\Gamma}{ \beta _2},\cr \\ \mu =-2\frac{\gamma }{%
\beta _2},\cr } 
\end{equation}

\noindent
we have : 

\begin{equation}
\frac{d^2F(\tau )}{d\tau ^2}+sgn(\lambda )|\lambda |F(\tau )+sgn(\mu )|\mu |{%
\ \frac{F(\tau )}{1+F^2(\tau )}}=0, 
\end{equation}

With a change of variables similar to Eqn. (11), Eqn. (26) reduces to: 

\begin{equation}
Q^{^{\prime 2}} +sgn(\lambda) Q + sgn(\mu) {\frac{Q}{a^2 +Q^2}}=0, 
\end{equation}

\noindent
where $a^{2} ={\frac{| \lambda | }{|\mu |}}$ $={\frac{|
\Gamma | }{| \gamma |}} $. Using Eqn.(22) and the condition for bright soliton \cite {m11}, $\Gamma$, soliton propagation constant is equal to the PR nonlinearity $\Delta K_{NL}$, i.e.   ${\frac{\Gamma}{\Delta K_{NL}}}=1 $ which further gives us the condition that $|\Gamma|/|\gamma| < 1 $ (since $I_s/I_p$ is always positive ). Similarly for dark solitons  we have ${\frac{\Gamma}{\Delta K_{NL}}}= {\frac{\ln (1+Q_0^{2})}{Q_0^{2}}}.$ Since ${\frac{\ln (1+Q_0^{2})}{Q_0^{2}}}<1$ for every value of $Q_0$, we have again  $|\Gamma|/|\gamma| < 1 $. 

Similar to last section, we multiply Eqn. (27) by $Q^{\prime}$ and integrate, we obtain the corresponding Hamiltonian function:

\begin{equation}
H(Q,Q^{\prime})=Q^{{\prime}2}+ sgn(\lambda) Q^2 + sgn(\mu)\ln(a^2
+Q^2)=h, 
\end{equation}

\noindent
where $h$ is a constant of integration.

We look in the phase space for solitary solutions of the above equation. Using Eqn.(15) we find as before the extrema points of the Hamiltonian to be:

\begin{equation}
\cases{ h = sgn(\lambda)[-\frac {sgn(\mu)}{sgn(\lambda)}-a^2] + sgn(\mu)\ln(-\frac{sgn(\mu)}{sgn(\lambda)}) \cr
              h= sgn(\mu) \ln(a^2) }
\end{equation}

\noindent
If $sgn(\lambda) = sgn(\mu)$ then only the second value of $h$ is valid.

 First we consider the case of $\mu<0 , \lambda>0$, where the two values of $h$ are 
$h = 1-a^2$ and $h = -\ln(a^2)$. It can be easily seen that there is no solitary solution for the 
first value of $h$. For the second value of $h$ Eqn. (28) can be expressed as: 

\begin{equation}
H(Q,Q^{\prime })=Q^{\prime 2}+ Q^2 -\ln(a^2 +Q^2)= - \ln a^2 
\end{equation}

The shapes of the orbits for different values of $a^2$ are shown in Fig.(5).
We obtain the following differential :

\begin{equation}
{\frac{d Q}{\sqrt {\ln(a^2 + Q^2) -Q^2 -\ln (a^2)}}}=\pm d T. 
\end{equation}

Integrating this equation, we find the  profile of the solution as plotted in Fig.(6). 

\noindent
For the case $\mu > 0,\lambda < 0$, the solitary wave solution  is located at $h=a^2-1$ and substituting $h$ in the Hamiltonian of Eqn.(28), we can derive a dark solitary solution. The phase space orbits are  given in Fig.(7) and the profile of the dark solution is given in Fig.(8).

\noindent
For all other cases ($\mu < 0, \lambda < 0$ and $\mu > 0, \lambda > 0$), there are no solitary solutions.

\section{Conclusions}

We started from the NLS equation and solved it for a time-dependent potential in a medium with Pockels effect.  Our analytical solutions show that there is only one possible case where the soliton-like wave can propagate in the Pockels material.

In the second part, we solved the NLS equation for PR medium to find the conditions under which the soliton-like waves can propagate. The conditions we found are: a) $\mu<0, \lambda>0$ that leads to the bright solitary waves and b)$\mu>0, \lambda<0$, which leads to the dark solitary waves.

\newpage

\section{Appendix}

Nonlinear index of refraction of the PR medium be derived from the two beam coupling (pump and signal) inside a PR material.  Nonlinear polarization that is produced by the coupling of two beams is given by \cite{m8}, \cite {m15}:

\begin{equation}
P^{NL}=(\frac {\Delta \epsilon}{4 \pi}e^{i {q\cdot r}} + c.c. )(A_s e^{ik_s \cdot r} + A_p e^{ik_p \cdot r})
\end{equation} 

\noindent
The part of nonlinear polarization that has the spatial variation $e^(ik_s.r)$ can act as a phase-matched source for the signal wave as:

\begin{equation}
p_s^{NL} = \frac{\Delta \epsilon ^*}{4 \pi} A_p e^{i k_s \cdot r} 
                = -i \frac{\epsilon^2 \gamma_{eff} E_m}{4 \pi} \frac{|A_p|^2 A_s}{|A_p|^2 + |A_s|^2} e^{i k_s \cdot r} 
\end{equation}

\noindent
Using the nonlinear relationship between $P^{NL}_s$ and $\epsilon ^{NL} $ and Eqn.(33), we can write
\begin{eqnarray}
4 \pi p_s^{NL} = \epsilon^{NL} E_s = \epsilon^{NL} A_s e^{i k_s \cdot r} = -i \epsilon^2 \gamma_{eff} E_m \frac{I_p}{I_p + I_s} A_s e^{i k_s \cdot r}
\end{eqnarray}

\noindent
as a reseult of which the value of $\epsilon ^{NL}$ is equal to:

\begin{equation}
\epsilon^{NL} = \epsilon^2 \gamma_{eff} E_m \frac{I_p}{I_p + I_s}.
\end{equation}

Now, given the value of $\epsilon ^{NL}$, we can calculate the $\Delta k_{NL}$ for the PR medium by substituting $\epsilon=n^2$ and $k_0=\frac{n\omega}{c}$ in the nonlinear relation $\Delta k_{NL}=\frac{1}{2k_0} \frac{\omega ^2}{c^2} \epsilon ^{NL}$ as follows:

\begin{equation}
\Delta k_{NL} = \gamma \frac{I_p}{I_p + I_s},
\end{equation}

\noindent
where we have defined a parameter $\gamma$ as:

\begin{equation}
\gamma = \frac{1}{2} \frac{\omega}{c} n^3 \gamma_{eff} E_m.
\end{equation}

\noindent
Eqn. (36) is the potential that we use in the (NLS) equation  for PR media in Eqn. (23) of the main text.

\section{Acknowledgement}
S.A. Hosseini acknowledges  Dr. Alok Sharan and Dr. Dey Rukmini for useful discussions.

\begin{figure}
\caption{ Phase space diagram of $Q^{\prime 2} -Q^2 +\frac {2}{3} Q^3 = h $ for different values of h. The curves 1,2,3 and 4 correspond to $h=0 $, 
$ -\frac {1}{3} < h < 0 (h=-\frac {1}{6})$, $h > 0 (h=\frac {1}{6})$ and $h=-\frac {1}{3}$, respectively.}

\caption{ Normalized amplitude profile of $|A(z,t)| = \frac{3 |\lambda |}{2 |\mu |}  sech^2 (\frac{\sqrt {|\lambda |}}{2} (t - z/v_g) )$ for $\lambda < 0 $ and $\mu > 0$, which shows bright solitary solution. } 

\caption{ Phase space diagram of $Q^{\prime 2} +Q^2 +\frac {2}{3} Q^3 = h $ for different value of h. The curves 1,2,3 and 4 correspond to $ h=\frac {1}{3}$, $ 0 < h < \frac {1}{3} (h=\frac {1}{6})$, $h \le 0 (h=0)$ and $h>\frac {1}{3}$, respectively.}

\caption{Normalized amplitude profile of $ |A(z,t)| = \frac{|\lambda|}{|\mu|} [ -\frac{3}{2} tanh^2 ( \frac{\sqrt{\lambda}}{2} (t -  z/v_g ) ) + \frac{1}{2} ]$ for $\lambda > 0$ and $\mu > 0$ in Pockels media. This profile shows that there is no solitary solution possible under this condition.}

\caption{ Phase space diagram of $Q^{\prime 2} +Q^2 -\ln(a^2 +Q^2) =  -\ln a^2 $ for different value of $a^2$. The curves 1,2,3 and 4 correspond to $ a^2= 0.2$, $ 0.4 $, $ 0.6 $ and $ 0.8$, respectively.}

\caption{Amplitude profile of the differential ${\frac{d Q}{\sqrt {\ln(a^2 + Q^2) -Q^2 -\ln (a^2)}}}=\pm d T $ solved numerically for $a^2=0.6 $, which shows the bright solitary solution.}

\caption{ Phase space diagram of $Q^{\prime 2} -Q^2 +\ln(a^2 +Q^2) = a^2-1 $ for different value of $a^2$. The curves 1,2,3 and 4 correspond to $ a^2= 0.2$, $ 0.4 $, $ 0.6 $ and $ 0.8$, respectively}

\caption{Amplitude profile of the differential ${\frac{d Q}{\sqrt {Q^2 -\ln(a^2 + Q^2) a^2 -1}}}=\pm d T $ solved numerically for $a^2=0.75 $, which shows the dark solitary solution.}
\end{figure}

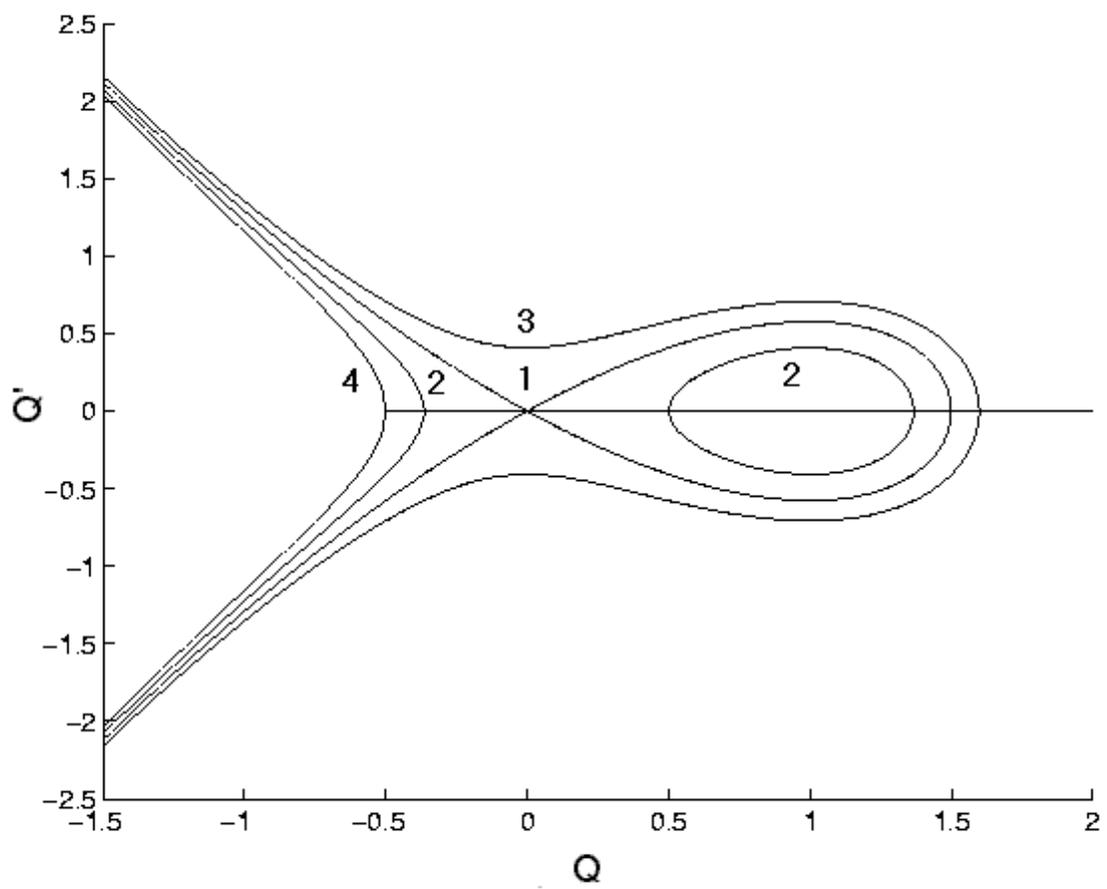
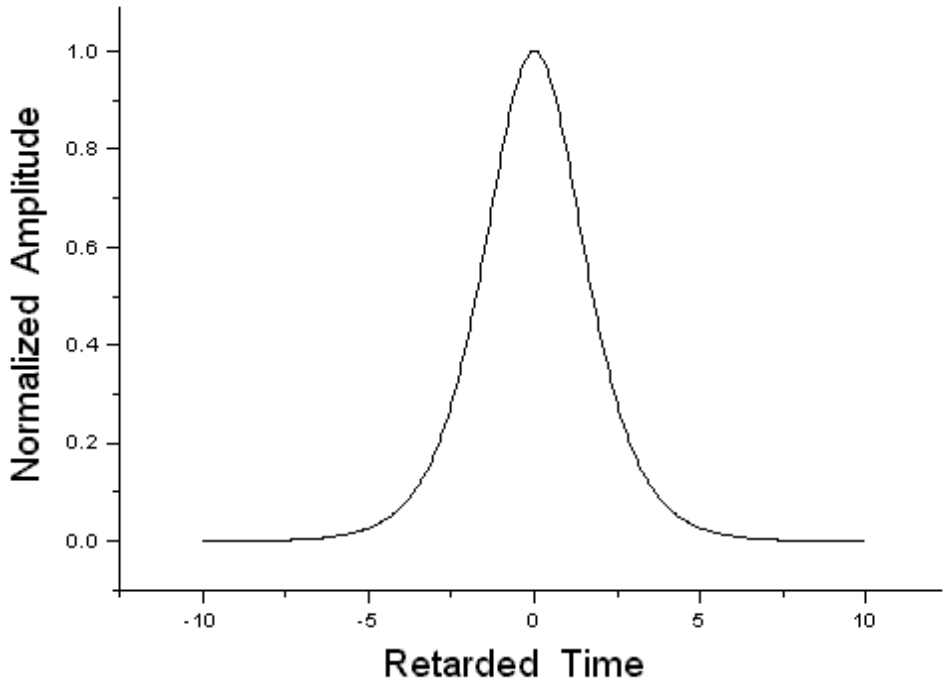
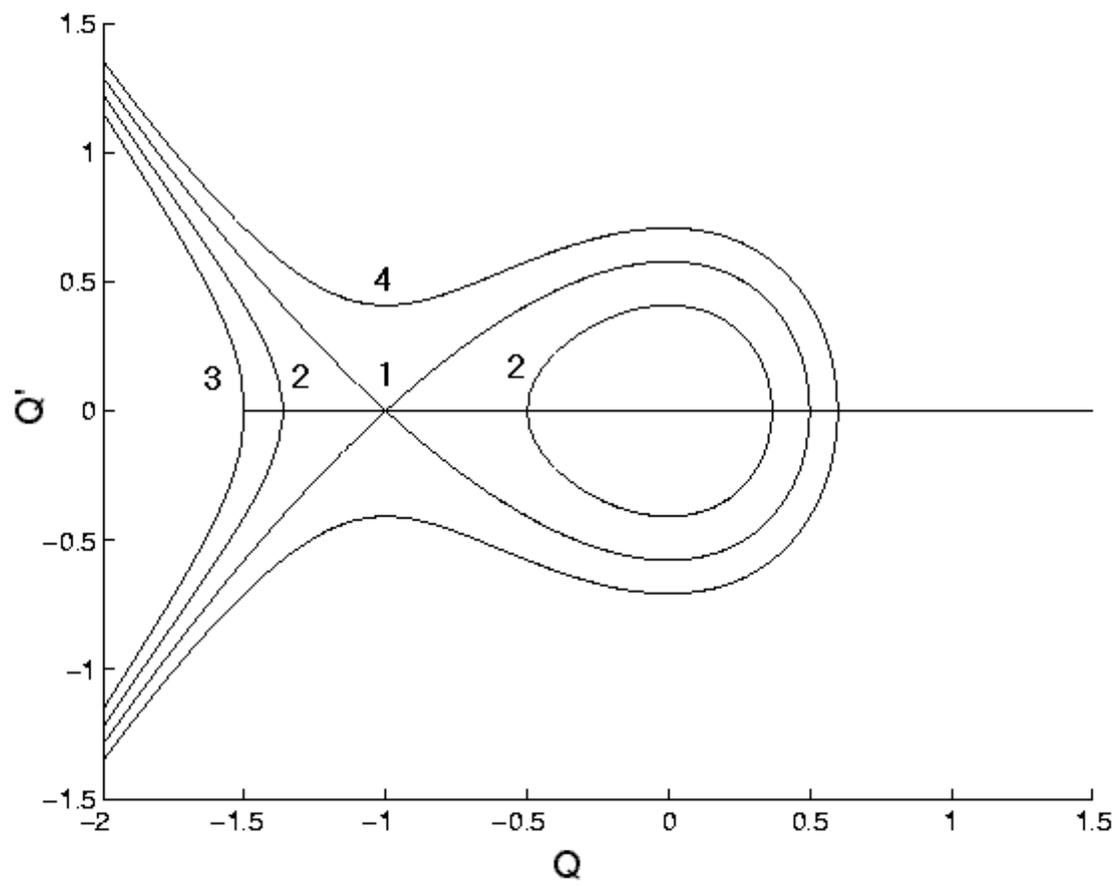
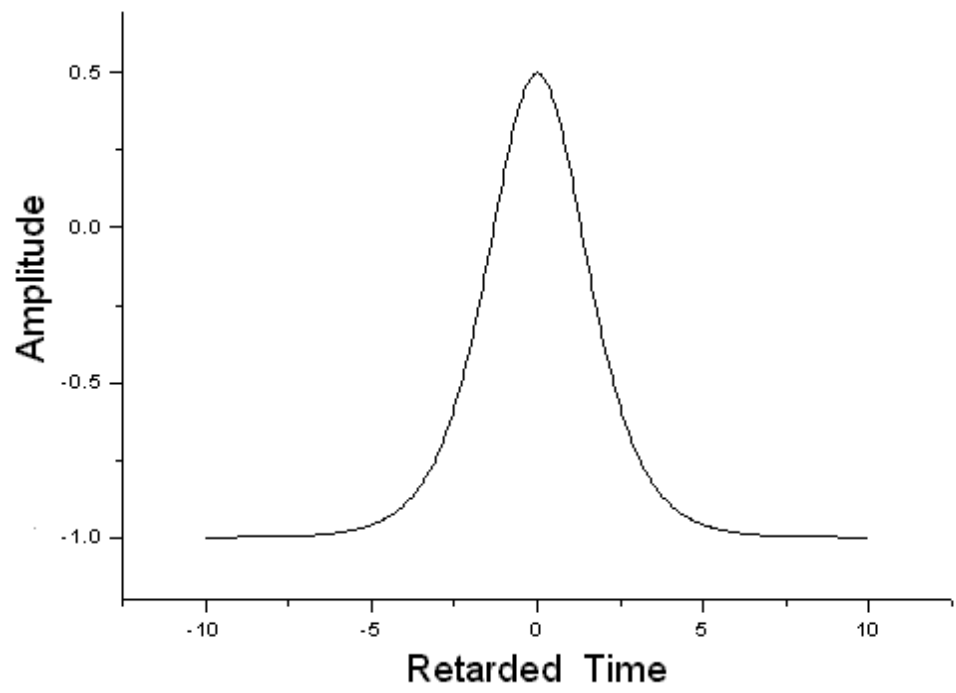
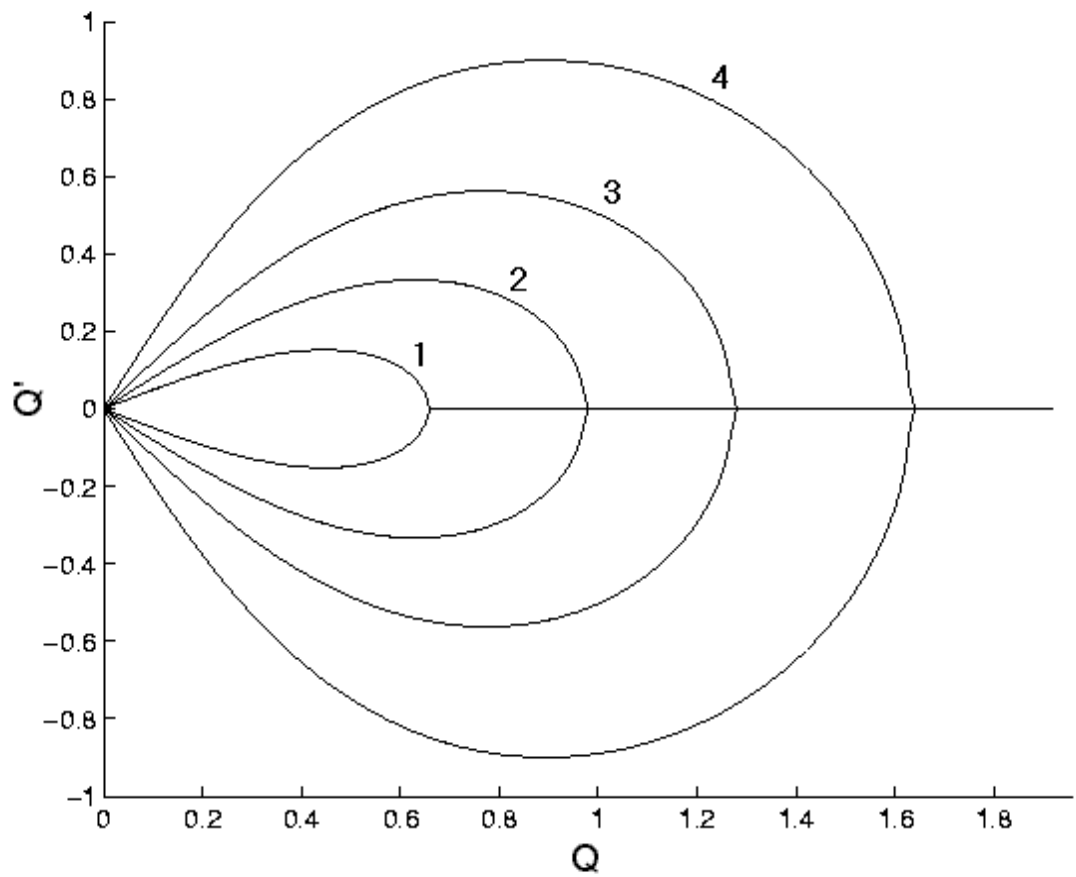
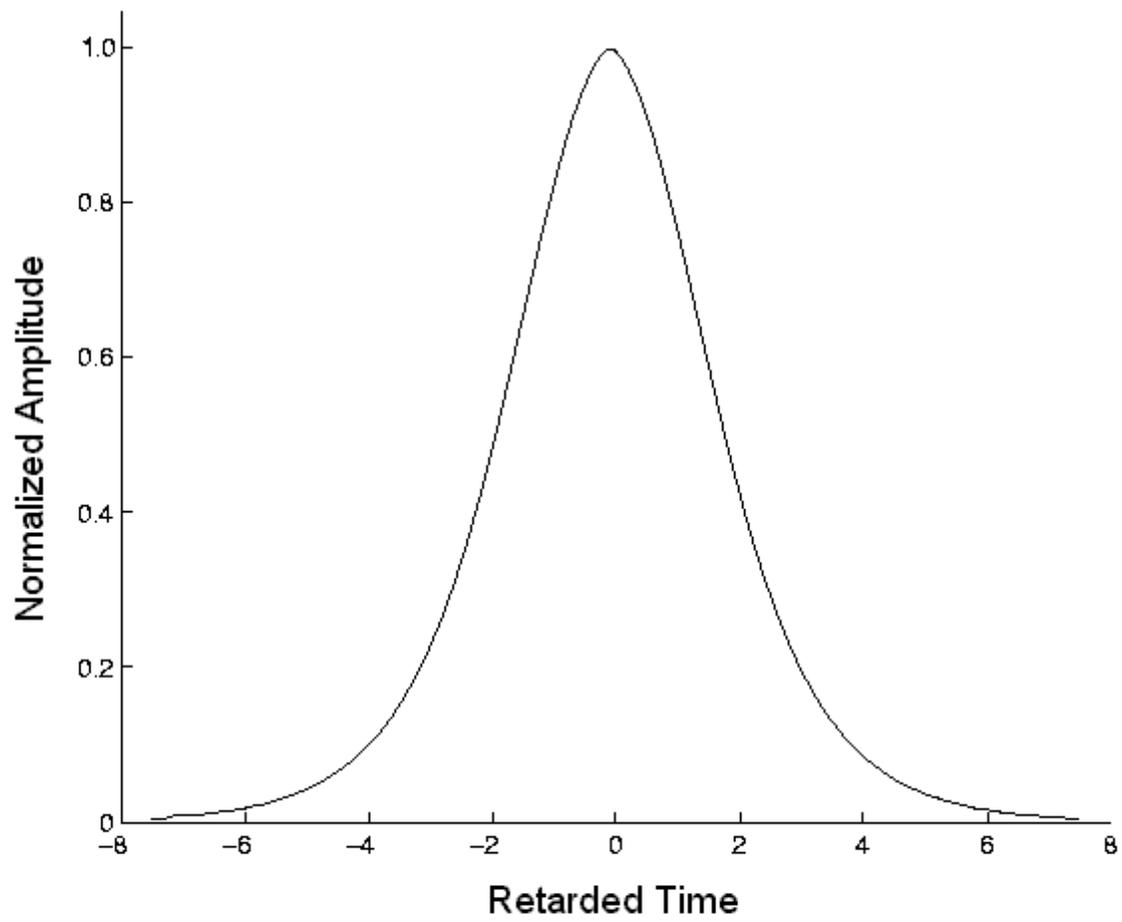
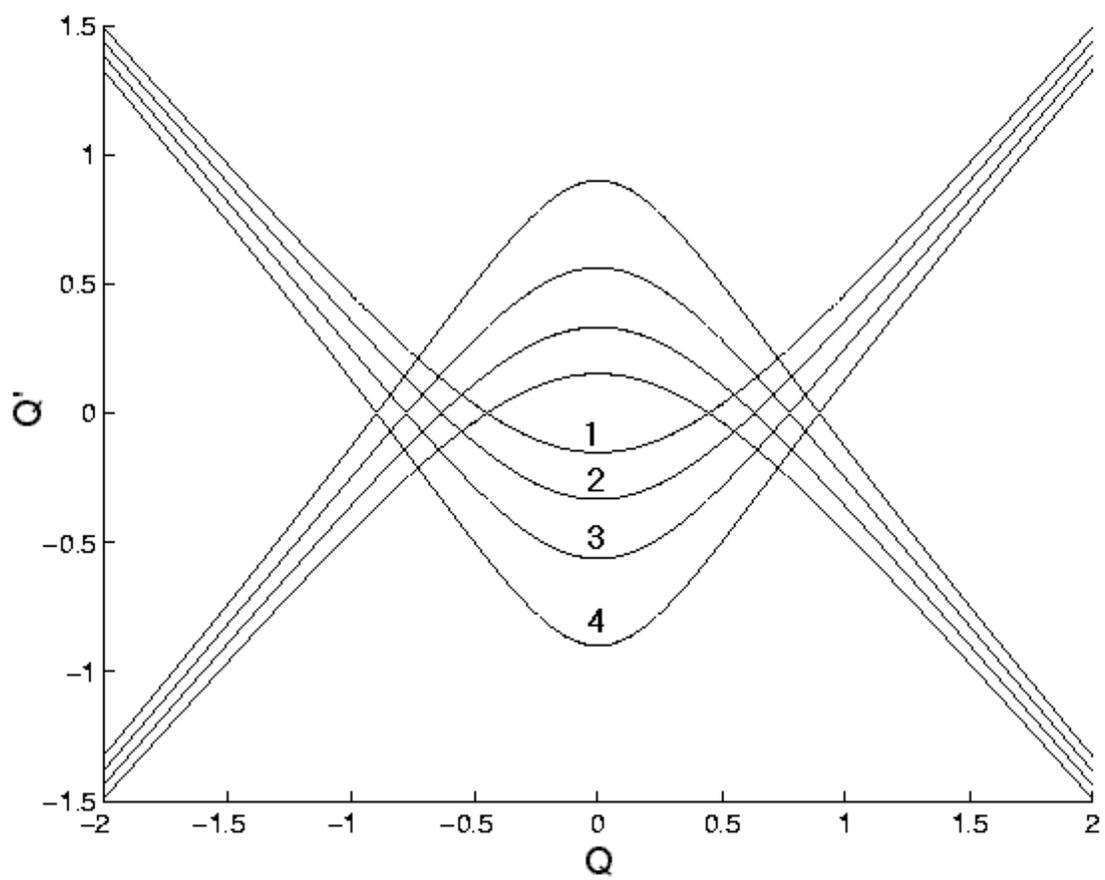
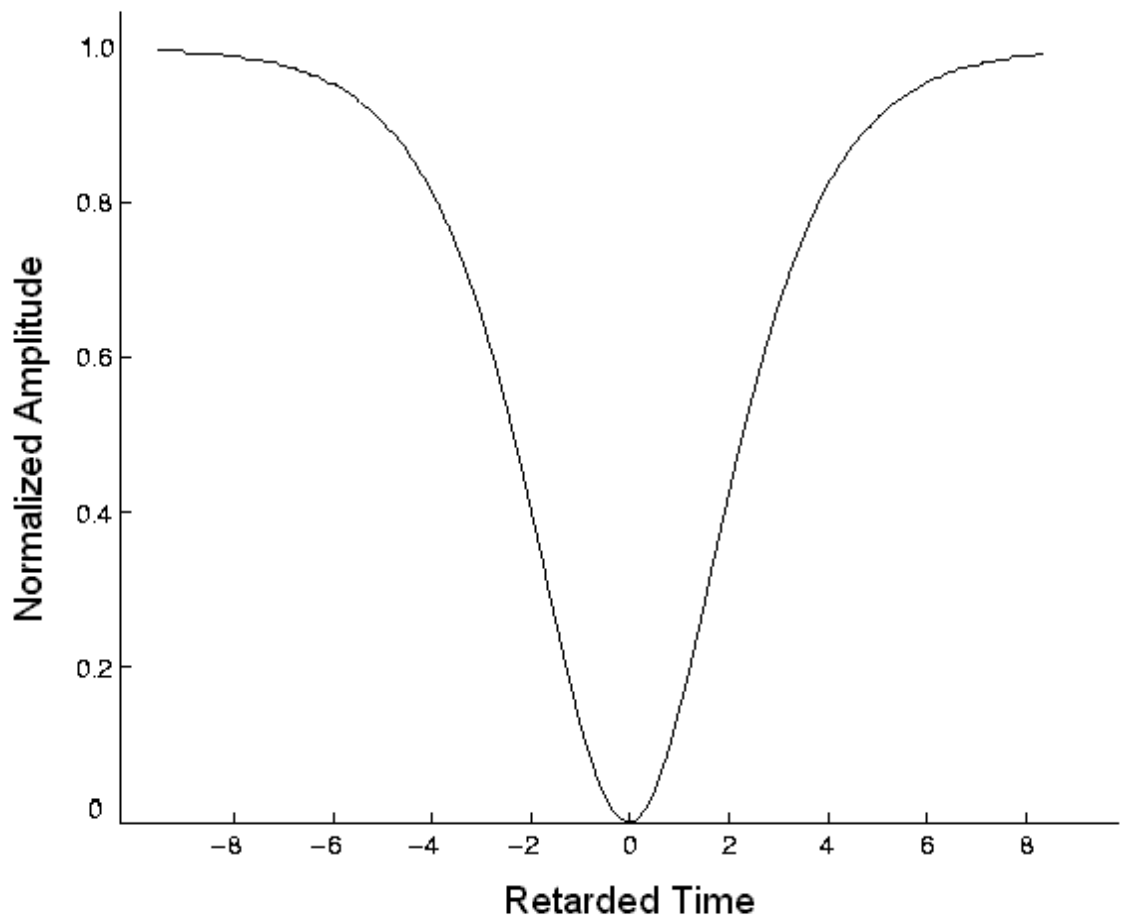
\end{document}